\documentclass[english,12pt]{article}
\usepackage[cp1251]{inputenc}
\usepackage{babel}
\usepackage{amsfonts, amsmath}

\newcommand{\be}{\begin{equation}} \newcommand{\ee}{\end{equation}}

\newcommand{\gsim}{\mathrel{\hbox{\rlap{\lower.55ex\hbox{$\sim$}} \kern-.3em \raise.4ex \hbox{$>$}}}}

\begin{document}
\begin{center}
{\bf Minimal Length,Minimal Inverse Temperature,Measurability and
Black Holes}\\
\vspace{5mm} Alexander Shalyt-Margolin \footnote{E-mail:
a.shalyt@mail.ru; alexm@hep.by}\\ \vspace{5mm} \textit{Research
Institute for Nuclear Problems,Belarusian State University, 11
Bobruiskaya str., Minsk 220040, Belarus}
\end{center}
PACS (2010): 04.60.-m; 98.80.Cq; 04.70.Dy
\\
 \rm\normalsize \vspace{0.5cm}

\begin{abstract}
The measurability notion introduced previously in a quantum theory
on the basis of a minimal length in this paper is defined in
thermodynamics on the basis of a minimal inverse temperature.
Based on this notion, some inferences are made for gravitational
thermodynamics of horizon spaces and, specifically, for black
holes with the Schwarzschild metric.
\end{abstract}

\section{Introduction.}

This paper is a continuation of the earlier works published
 by the author \cite{Shalyt-AHEP2},\cite{Shalyt-Entropy2016.1}.
The main idea and target of these works is to construct a correct
quantum theory and gravity in terms of the variations (increments)
dependent on the existent energies.
\\It is clear that such a theory should not involve infinitesimal space-time variations
\begin{equation}\label{Introd.1}
dt,dx_{i},i=1,...,3.
\end{equation}
Besides, as shown in \cite{Shalyt-Entropy2016.1}, with the
involvement of some universal units of the minimal length
$l_{min}$  and time $t_{min}$, this theory   will be discrete and
dependent on the explicitly defined discrete parameters but very
close to the initial continuous theory at low energies.
\\The notion of {\bf measurability} introduced by the author in \cite{Shalyt-Entropy2016.1}
is essential for the above-mentioned discrete theory.
\\This paper demonstrates that an analogous notion may be introduced
in thermodynamics on the basis of the minimal inverse temperature.
Based on the obtained results, the inferences for gravitational
thermodynamics of horizon spaces and, specifically, for black
holes with the Schwarzschild metric are introduced.
\\ All aspects of the author’s motivation were given in detail
in \cite{Shalyt-AHEP2} and, in particular, in \cite{Shalyt-Entropy2016.1}.
\\ In short, the motivation is as follows.
\\ According to the present-day views, both quantum theory and gravity in the ultraviolet region
is related to some new parameters defined at high (apparently
Planck) energies (for example \cite{QG1}). But at low energies,
this relationship is not apparent due to its insignificant effect
in this case, on the one hand, and due to the mathematical
apparatus of continuous space-time, where the existent theories
are considered, on the other hand.
\\ By the author's opinion, the correct definition of a dynamics
of the above-mentioned relationship at all the energy scales will
enable us to find the key to solutions of all the problems given
below:
\\
{\bf 1.1} ultraviolet and infra-red divergences in a quantum field theory;
\\
{\bf 1.2} correct transition to the high-energy (quantum) region for gravity;
\\{\bf 1.3} and possibility of the existence of ''nonphysical'' solutions
for the General Relativity (for example, the solutions involving
the {\bf Closed Time-like Curves} (CTC)
\cite{Godel}--\cite{Lobo}).
\\To make this paper maximally self-contained, the author includes
all the required earlier obtained results precisely with the
corresponding references in Subsection 2.1 and at the beginning of
Subsection 2.2.
\\New results are presented in the second half of Subsection
2.2., and also in Subsection 2.3. and in Section 3.

\section{Minimal Length, Minimal Inverse Temperature, and Measurability}

\subsection{Generalized Uncertainty Principles in Quantum Theory
and Thermodynamics}

In this Subsection the author presents some of the results from Section
2 of the paper \cite{shalyt3},because they are important for this work.
\\It is well known that in thermodynamics an inequality for the pair
interior energy - inverse temperature that is completely analogous
to the standard uncertainty
 relation in quantum mechanics \cite{Heis1} can be written \cite{Lind} -- \cite{Lind5}. The
 only (but essential) difference of this inequality from the quantum mechanical
 one is that the main quadratic fluctuation is defined by means of
 the classical partition function rather than by the quantum mechanical expectation values.
 In the last years a lot of papers appeared in which the usual
 momentum-coordinate uncertainty relation has been modified at very high
 energies on the order of the Planck energy $E_p$ \cite{Ven1}--\cite{Nozari}. In this note we
 propose simple reasons for modifying the thermodynamic uncertainty relation at
 Planck energies. This modification results in existence of the minimal
 possible main quadratic fluctuation of the inverse temperature. Of course we
 assume that all the thermodynamic quantities used are properly defined so that
 they have physical sense at such high energies.

We start with usual Heisenberg Uncertainty Principle (relation)
\cite{Heis1} for momentum - coordinate:
\begin{equation}\label{U1}
 \Delta x\geq\frac{\hbar}{\Delta p}.
\end{equation}
 It was shown that at the Planck
 scale a high-energy term must appear:
\begin{equation}\label{U2}
\Delta x\geq\frac{\hbar}{\Delta p}+ \alpha^{\prime}
l_{p}^2\frac{\triangle p}{\hbar}
\end{equation}
where $l_{p}$ is the Planck length $l_{p}^2 = G\hbar /c^3 \simeq
1,6\;10^{-35}m$ and $\alpha^{\prime}$ is a constant. In
\cite{Ven1} this term is derived from the string theory, in
\cite{GUPg1}
 it follows from the simple estimates of Newtonian gravity and quantum mechanics,
 in \cite{Magg2} it comes from the black hole physics, other methods can also be
 used \cite{Magg1},\cite{Magg3},\cite{Capoz}.
Relation (\ref{U2}) is quadratic in $\Delta p$
\begin{equation}\label{U4}
\alpha^{\prime} l_{p}^2\, ({\Delta p})^2 - \hbar\,\Delta x\Delta p
+ \hbar^2 \leq0
\end{equation}
 and therefore leads to the fundamental length
\begin{equation}\label{U5}
 \Delta x_{min}=2 \surd \alpha^{\prime} l_{p}
\end{equation}
Inequality (\ref{U2}) is called the Generalized Uncertainty
  Principle (GUP) in Quantum Theory.
 \\Using relations (\ref{U2}) it is easy to obtain a similar relation for the
 energy - time pair. Indeed (\ref{U2}) gives
\begin{equation}\label{U6}
\frac{\Delta x}{c}\geq\frac{\hbar}{\Delta p c }+\alpha^{\prime}
l_{p}^2\,\frac{\Delta p}{c \hbar},
\end{equation}
then
\begin{equation}\label{U7}
\Delta t\geq\frac{\hbar}{\Delta
E}+\alpha^{\prime}\frac{l_{p}^2}{c^2}\,\frac{\Delta p
c}{\hbar}=\frac{\hbar}{\Delta
E}+\alpha^{\prime}t_{p}^2\,\frac{\Delta E}{ \hbar},
\end{equation}
where the smallness of $l_p$ is taken into account so that the
difference between $\Delta E$ and $\Delta (pc)$ can be neglected
and $t_{p}$  is the Planck time
$t_{p}=l_p/c=\sqrt{G\hbar/c^5}\simeq 0,54\;10^{-43}sec$.
Inequality (\ref{U7}) gives analogously to (\ref{U2}) the lower
boundary for time $\Delta t\geq2t_{p}$, determining the
fundamental time
\begin{equation}\label{U10b}
 t_{min}=2\sqrt{\alpha^{\prime}}t_{p}.
 \end{equation}
 Thus, the inequalities discussed can be rewritten in a standard form
\begin{equation}\label{U11b}
\left\{ \begin{array}{ll} \Delta x &
\geq\frac{\displaystyle\hbar}{\displaystyle\Delta
p}+\alpha^{\prime} \left(\frac{\displaystyle\Delta
p}{\displaystyle P_{pl}}\right)\,
\frac{\displaystyle\hbar}{\displaystyle P_{pl}}
\\
 & \\
 \Delta t & \geq\frac{\displaystyle\hbar}{\displaystyle\Delta E}+\alpha^{\prime}
 \left(\frac{\displaystyle\Delta E}{\displaystyle E_{p}}\right)\,
 \frac{\displaystyle\hbar}{\displaystyle E_{p}}
\end{array} \right.
\end{equation}
where $P_{pl}=E_p/c=\sqrt{\hbar c^3/G}$.
 Now we
consider the thermodynamic uncertainty relations between the
inverse temperature and interior energy of a macroscopic ensemble
\begin{equation}\label{U12}
\Delta \frac{1}{T}\geq\frac{k_{B}}{\Delta U},
\end{equation}
where $k_{B}$ is the Boltzmann constant.
\\N.Bohr \cite{r10} and
W.Heisenberg \cite{Heis-term} first pointed out that such kind of
uncertainty principle should take place in thermodynamics. The
thermodynamic uncertainty  relations (\ref{U12})  were proved by
many authors and in various ways \cite{Lind} -- \cite{Lind5}.
Therefore their validity does raises no doubts. Nevertheless,
relation (\ref{U12}) was proved in view of the standard model of
the infinite-capacity heat bath encompassing the ensemble. But it
is obvious from the above inequalities that at very high energies
the capacity of the heat bath can no longer be assumed infinite at
the Planck scale. Indeed, the total energy of the pair heat bath -
ensemble may be arbitrary large but finite, merely as the Universe
was born at a finite energy. Hence the quantity that can be
interpreted as the a temperature of the ensemble must have the
upper limit and so does its main quadratic deviation. In other
words, the quantity $\Delta (1/T)$ must be bounded from below. But
in this case an additional term should be introduced into
(\ref{U12})
\begin{equation}\label{U12a}
\Delta \frac{1}{T}\geq\frac{k_{B}}{\Delta U} + \eta\,\Delta U
\end{equation}
where $\eta$ is a coefficient. Dimension and symmetry reasons give
\begin{equation}\label{U12a*}
\eta \sim \frac{k_{B}}{E_p^2}\enskip or\enskip \eta =
\alpha^{\prime} \frac{k_{B}}{E_p^2}
\end{equation}
 As in the previous cases,
inequality (\ref{U12a}) leads to the fundamental (inverse)
temperature
\begin{eqnarray}\label{U15}
T_{max}=\frac{\hbar}{2\surd \alpha^{\prime}t_{p}
k_{B}}=\frac{E_p}{2\surd \alpha^{\prime} k_{B}}=\frac{T_p}{2\surd
\alpha^{\prime}
}=\frac{\hbar}{t_{min} k_{B}},\nonumber\\
 \quad \beta_{min} = {1\over
k_{B}T_{max}} =  \frac{t_{min}}{\hbar}.
\end{eqnarray}
In the work \cite{Farmany}   the black hole horizon temperature
has been measured with the use of the Gedanken experiment. In the
process the Generalized Uncertainty Relations in Thermodynamics
(\ref{U12a}) have been derived also. Expression (\ref{U12a}) has
been considered in the monograph \cite{Carroll} within the scope
of the mathematical physics methods.
\\Thus,we obtain a
system of the generalized uncertainty relations in the symmetric
form
\begin{equation}\label{U17}
\left\{
\begin{array}{lll}
\Delta x & \geq & \frac{\displaystyle\hbar}{\displaystyle\Delta
p}+ \alpha^{\prime} \left(\frac{\displaystyle\Delta
p}{\displaystyle P_{pl}}\right)\,
\frac{\displaystyle\hbar}{\displaystyle P_{pl}}+... \\ &  &  \\
\Delta t & \geq & \frac{\displaystyle\hbar}{\displaystyle\Delta
E}+\alpha^{\prime} \left(\frac{\displaystyle\Delta
E}{\displaystyle E_{p}}\right)\,
\frac{\displaystyle\hbar}{\displaystyle E_{p}}+...\\
  &  &  \\
  \Delta \frac{\displaystyle 1}{\displaystyle T}& \geq &
  \frac{\displaystyle k_{B}}{\displaystyle\Delta U}+\alpha^{\prime}
  \left(\frac{\displaystyle\Delta U}{\displaystyle E_{p}}\right)\,
  \frac{\displaystyle k_{B}}{\displaystyle E_{p}}+...
\end{array} \right.
\end{equation}
or in the equivalent form
\begin{equation}\label{U18}
\left\{
\begin{array}{lll}
\Delta x & \geq & \frac{\displaystyle\hbar}{\displaystyle\Delta
p}+\alpha^{\prime} l_{p}^2\,\frac{\displaystyle\Delta
p}{\displaystyle\hbar}+... \\
  &  &  \\
  \Delta t & \geq &  \frac{\displaystyle\hbar}{\displaystyle\Delta E}+\alpha^{\prime}
  t_{p}^2\,\frac{\displaystyle\Delta E}{\displaystyle\hbar}+... \\
  &  &  \\

  \Delta \frac{\displaystyle 1}{\displaystyle T} & \geq &
  \frac{\displaystyle k_{B}}{\displaystyle\Delta U}+\alpha^{\prime}
  \frac{\displaystyle 1}{\displaystyle T_{p}^2}\,
  \frac{\displaystyle\Delta U}{\displaystyle k_{B}}+...,
\end{array} \right.
\end{equation}
where the dots mean the existence of higher order corrections as
in \cite{r21}.
 Here $T_{p}$ is the Planck temperature:
$T_{p}=E_{p}/k_{B}$.
\\In literature the relation (\ref{U12}) is
referred to as the Uncertainty Principle in Thermodynamics (UPT).
Let us call the relation (\ref{U12a}) the Generalized Uncertainty
Principle in Thermodynamics (GUPT).
\\In this case, without the loss of generality and for symmetry,
it is assumed that a dimensionless constant in the right-hand side
of GUP (formula (\ref{U2})) and in the right-hand side of GUPT
(formula (\ref{U12a})) is the same -- $\alpha^{\prime}$.

\subsection{Minimal Length and Measurability Notion in Quantum Theory}

First, we consider in this Subsection the principal definitions
from \cite{Shalyt-AHEP2},\cite{Shalyt-Entropy2016.1}which are
required to derive the key formulae in the second part of the
Subsection and to obtain further results.
\\
\\{\bf Definition I.} Let us call as {\bf primarily measurable
variation} any small variation (increment) $\widetilde{\Delta}
x_{\mu}$ of any spatial coordinate $x_{\mu}$ of the arbitrary
point $x_{\mu},\mu~=~1,...,3$ in some space-time system $\emph{R}$
if it may be realized in the form of the uncertainty (standard
deviation) $\Delta x_{\mu}$ when this coordinate is measured
within the scope of Heisenberg's Uncertainty Principle (HUP)
 \cite{Heis1} (formula (\ref{U1}) in the general case):
\begin{equation}\label{HUP01}
\widetilde{\Delta} x_{\mu}=\Delta x_{\mu},\Delta x_{\mu}\simeq
\frac{\hbar}{\Delta p_{\mu}}, \mu=1,2,3
\end{equation}
for some $\Delta p_{\mu}\neq 0$.
\\Similarly, at $\mu=0$ for pair ``time-energy'' $(t,E)$, let us call
any small variation (increment) the {\bf primarily measurable
variation}  in the value of time $\widetilde{\Delta}
x_{0}=\widetilde{\Delta} t_{0}$ if it may be realized in the form
of the uncertainty (standard deviation) $\Delta x_{0}=\Delta t$
and~then
\begin{equation}\label{HUP02}
\widetilde{\Delta} t=\Delta t,\Delta t\simeq \frac{\hbar}{\Delta
E}
\end{equation}
for some $\Delta E\neq 0$.  Formula (\ref{HUP02}) is nothing else
but as  formula (\ref{U7}) for $\Delta E\ll E_p$.
\\Here HUP is given for the nonrelativistic case. In
the relativistic case HUP has the distinctive features
\cite{Land2} which, however, are of no significance for the
general formulation of {\bf Definition I.},being associated only
with particular alterations in the right-hand side of the second
relation Equation (\ref{HUP02}).
\\It is clear that at low energies $E\ll E_P$ (momenta $P\ll
P_{pl}$) {\bf Definition I.} sets a lower bound for the {\bf
primarily measurable variation}  $\widetilde{\Delta} x_{\mu}$ of
any space-time coordinate $x_{\mu}$.
\\At high energies $E$ (momenta $P$) this is not the case if $E$ ($P$) has no upper
limit. But, according to the modern knowledge, $E$ ($P$) is
bounded by some maximal quantities $E_{max}$, ($P_{max}$)
\begin{equation}\label{HUP03}
E\leq E_{max},P\leq P_{max},
\end{equation}
where,in general,  $E_{max},P_{max}$  may be on the order of the
Planck quantities $E_{max}\propto E_P,P_{max}\propto P_{pl}$ and
also may be the trans-Planck's quantities.

In any case the quantities $P_{max}$  and $E_{max}$ lead to the
introduction of the minimal length $l_{min}$ and of the minimal
time $t_{min}$.
\\{\bf Supposition II.} There is the minimal length
$l_{min}$  as {\it a minimal measurement unit} for all {\bf
primarily measurable variations} having the dimension of length,
whereas the minimal time $t_{min}=l_{min}/c$ as {\it a minimal
measurement unit} for all quantities or {\bf primarily measurable
variations (increments)}  having the dimension of time, where $c$
is the speed of light.

$l_{min}$ and $t_{min}$ are naturally introduced as $\Delta
x_{\mu},\mu=1,2,3$  and $\Delta t$  in Equations (\ref{HUP01}) and
(\ref{HUP02}) for $\Delta p_{\mu}=P_{max}$ and $\Delta E=E_{max}$.

For definiteness, we consider that $E_{max}$ and $P_{max}$ are the
quantities on the order of the Planck quantities, then $l_{min}$
and $t_{min}$ are also on the order of the Planck quantities
$l_{min}\propto l_P$, $t_{min}\propto t_P$.

{\bf Definition I.} and {\bf Supposition II.} are quite natural in
the sense that there  are no physical principles with which they
are inconsistent.
\\The combination of {\bf Definition I.} and
{\bf Supposition II.} will be called the {\bf Principle of Bounded
Primarily Measurable Space-Time Variations (Increments)} or for
short {\bf Principle of Bounded  Space-Time Variations
(Increments)} with abbreviation (PBSTV).
\\As the minimal unit of measurement $l_{min}$ is available
 for all the {\bf primarily measurable variations} $\Delta L$
 having the dimensions of length, the
{``Integrality Condition'' (IC)} is the case
 \begin{equation}\label{Introd 2.4}
 \Delta L=N_{\Delta L}l_{min},
 \end{equation}
where~$N_{\Delta L}~>~0$~is an integer number.
\\In a like manner, the same {``Integrality Condition'' (IC)}
is the case for all the {\bf primarily measurable variations}
$\Delta t$ having the dimensions of time. And similar to Equation
(\ref{Introd 2.4}), we get the following expression for any time
$\Delta t$:
 \begin{equation}\label{Introd 2.4new}
\Delta t\equiv \Delta t(N_{t})=N_{\Delta t}t_{min},
\end{equation}
where similarly $N_{\Delta t}~>~0$ is an integer number too.
\\
{\bf Definition 1} ({\textbf{Primary or Elementary
Measurability.}})
\\(1) {\it In accordance with the PBSTV, let us define
the quantity having the dimensions of length or time  as {\bf
primarily (or elementarily) measurable}, when it satisfies the
relation Equation (\ref{Introd 2.4}) (and respectively Equation
(\ref{Introd 2.4new})}).
\\(2){\it Let us define any physical quantity {\bf
primarily (or elementarily) measurable},
 when its value is consistent with points (1)  of this
 Definition.}
\\It is convenient to use the deformation parameter $\alpha_{a}$.
 This parameter has been introduced earlier in the papers
\cite{shalyt2},\cite{shalyt3},\cite{shalyt4}--\cite{shalyt7} as a
{\it deformation parameter} (in terms of paper \cite{Fadd}) on
going from the canonical quantum mechanics to  the quantum
mechanics at Planck's scales (Early Universe) that is considered
to be the quantum mechanics with the minimal length (QMML):
\begin{equation}\label{D1}
\alpha_{a}=l_{min}^{2}/a^{2},
\end{equation}
where $a$ is the measuring scale. It is easily seen that the
parameter $\alpha_{a}$  from Equation (\ref{D1}) is discrete as it
is nothing else but
\begin{equation}\label{D1.new1}
\alpha_{a}=l_{min}^{2}/a^{2}=\frac{l_{min}^{2}}{N^{2}_{a}l_{min}^{2}}=\frac{1}{N^{2}_{a}}.
\end{equation}
At the same time, from Equation (\ref{D1.new1}) it is evident that
$\alpha_{a}$ is irregularly discrete.
\\It should be noted that physical quantities complying with {\bf Definition 1}
are inadequate  for the research of physical systems.
 \\Indeed, such a variable as
\begin{eqnarray}\label{Def2}
\alpha_{N_{a}l_{min}}(N_{a}l_{min})=p(N_{a})\frac{l^{2}_{min}}{\hbar}
=l_{min}/N_{a},
\end{eqnarray}
where $\alpha_{N_{a}l_{min}}=\alpha_{a}$ is taken from formula
(\ref{D1.new1} at $a=N_{a}l_{min}$, and
$p(N_{a})=\frac{\hbar}{N_{a}l_{min}}$ is  the corresponding {\bf
primarily  measurable} momentum), is fully expressed in terms of
{\it only} {\bf Primarily Measurable Quantities} of {\bf
Definition 1} and  that's why it hence may appear at any stage of
calculations, but apparently
 does not complying with {\bf Definition 1}.
Because of this it is necessary to introduce the following
definition generalizing {\bf Definition 1}:
\\
\\{\bf Definition 2. Generalized Measurability}
\\We shall call any physical quantity as {\bf
generalized-measurable}  or  for  simplicity {\bf measurable} if
any of its values may be obtained in terms of {\bf Primarily
Measurable Quantities} of {\bf Definition 1}.
\\
\\In what follows, for simplicity, we will use the term {\bf
 Measurability} instead of {\bf Generalized Measurability}.
\\It is evident that any {\bf primarily measurable quantity (PMQ)} is {\bf measurable}.
Generally speaking, the contrary is not correct as indicated by
formula (\ref{Def2}).
\\The {\bf generalized-measurable} quantities  follow from
the {\bf Generalized Uncertainty Principle (GUP)} (formula
(\ref{U2})) that naturally leads to the minimal length $l_{min}$
\cite{Ven1}--\cite{Nozari}:
\begin{equation}\label{GUP1}
\Delta x_{min}=2 \surd \alpha^{\prime} l_{p}\doteq l_{min},
\end{equation}
For convenience, we denote the minimal length $l_{min}\neq 0$ by
$\ell$ and $t_{min}\neq 0$ by $\tau=\ell/c$.
\\ Solving inequality (\ref{U2}), in the case of equality we obtain the apparent formula
\begin{equation}\label{root1}
\Delta p_{\pm}=\frac{(\Delta x\pm \sqrt{(\Delta
x)^{2}-4\alpha^{\prime}l^{2}_{p}})\hbar}{2\alpha^{\prime}l^{2}_{p}}.
\end{equation}
Next, into this formula we substitute the right-hand part of
formula (\ref{Introd 2.4}) for $L=x$. Considering (\ref{GUP1}), we
can derive the following:
\begin{eqnarray}\label{root2}
\Delta p_{\pm}=\frac{(N_{\Delta x}\pm \sqrt{(N_{\Delta
x})^{2}-1})\hbar\ell}{\frac{1}{2}\ell^{2}}=\nonumber\\
=\frac{2(N_{\Delta x}\pm \sqrt{(N_{\Delta x})^{2}-1})\hbar}{\ell}.
\end{eqnarray}
But it is evident that at low energies $E\ll E_p;N_{\Delta x}\gg
1$ the plus sign  in the nominator (\ref{root2}) leads to the
contradiction as   it results in very high (much greater than the
Planck) values of $\Delta p$. Because of this, it is necessary to
select the minus sign in the numerator (\ref{root2}). Then,
multiplying the left and right sides of (\ref{root2}) by the same
number $N_{\Delta x}+ \sqrt{N_{\Delta x}^{2}-1}$, we get
\begin{eqnarray}\label{root3}
\Delta p=\frac{2\hbar}{(N_{\Delta x}+ \sqrt{N_{\Delta
x}^{2}-1})\ell}.
\end{eqnarray}
$\Delta p$ from formula (\ref{root3}) is the {\bf
generalized-measurable} quantity in the sense of {\bf Definition
2.} However, it is clear that at low energies $E\ll E_p$, i.e. for
$N_{\Delta x}\gg 1$, we have $\sqrt{N_{\Delta x}^{2}-1}\approx
N_{\Delta x}$. Moreover, we have
\begin{eqnarray}\label{root3.1}
\lim\limits_{N_{\Delta x}\rightarrow \infty}\sqrt{N_{\Delta
x}^{2}-1}=N_{\Delta x}.
\end{eqnarray}
Therefore, in this case (\ref{root3}) may be written as follows:
\begin{eqnarray}\label{root3.2}
\Delta p\doteq \Delta p(N_{\Delta x},
HUP)=\frac{\hbar}{1/2(N_{\Delta x}+ \sqrt{N_{\Delta
x}^{2}-1})\ell}\approx \frac{\hbar}{N_{\Delta
x}\ell}=\frac{\hbar}{\Delta x};N_{\Delta x}\gg 1,
\end{eqnarray}
in complete conformity with HUP. Besides, $\Delta p\doteq
\Delta p(N_{\Delta x},HUP)$, to a high accuracy,
 is a {\textbf{primarily measurable} quantity in the sense of {\bf
Definition 1}.
\\And vice versa it is obvious that at high energies $E\approx E_p$,
i.e. for $N_{\Delta x}\approx 1$, there is no way to transform
formula  (\ref{root3}) and we can write
\begin{eqnarray}\label{root3.3}
\Delta p\doteq \Delta p(N_{\Delta x},
GUP)=\frac{\hbar}{1/2(N_{\Delta x}+ \sqrt{N_{\Delta
x}^{2}-1})\ell};N_{\Delta x}\approx 1.
\end{eqnarray}
At the same time, $\Delta p\doteq \Delta p(N_{\Delta x},GUP)$ is a
{\textbf{Generalized Measurable} quantity in the sense of {\bf
Definition 2}.
\\ Thus, we have
\begin{equation}\label{root4}
GUP\rightarrow HUP
\end{equation}
for
\begin{equation}\label{root5}
 (N_{\Delta x}\approx 1)\rightarrow (N_{\Delta x}\gg 1).
\end{equation}
Also, we have
\begin{equation}\label{root6}
\Delta p(N_{\Delta x}, GUP)\rightarrow \Delta p(N_{\Delta x},HUP),
\end{equation}
where $\Delta p(N_{\Delta x}, GUP)$ is taken from formula
(\ref{root3.3}), whereas $\Delta p(N_{\Delta x}, HUP)$ -- from
formula (\ref{root3.2}).
\\
\\{\it Comment 2*.}
\\{\it From the above formulae it follows that, within GUP, the {\bf primarily measurable}
variations (quantities) are derived to a high accuracy from the
{\bf generalized-measurable} variations (quantities) {\it only} in
the low-energy limit $E\ll E_P$}
\\
\\ Next, within the scope of GUP, we can correct a value of the parameter
$\alpha_{a}$  from formula  (\ref{D1.new1}) substituting $a$ for
$\Delta x$    in the expression $1/2(N_{\Delta x}+ \sqrt{N_{\Delta
x}^{2}-1})\ell$.
\\Then at low energies $E\ll E_p$ we have the
{\textbf{primarily measurable} quantity $\alpha_{a}(HUP)$
\begin{eqnarray}\label{root3.2.}
\alpha_{a}\doteq \alpha_{a}(HUP)=\frac{1}{[1/2(N_{a}+
\sqrt{N_{a}^{2}-1})]^{2}}\approx \frac{1}{N^{2}_{a}};N_{a}\gg 1,
\end{eqnarray}
that corresponds, to a high accuracy, to the value from formula (\ref{D1.new1}).
\\ Accordingly, at high energies we have
$E\approx E_p$
\begin{eqnarray}\label{root3.3.}
\alpha_{a}\doteq \alpha_{a}(GUP)=\frac{1}{[1/2(N_{a}+
\sqrt{N_{a}^{2}-1})]^{2}};N_{a}\approx 1.
\end{eqnarray}
When going from high energies $E\approx E_p$ to low energies
$E\ll E_p$, we can write
\begin{eqnarray}\label{P3}
\alpha_{a}(GUP) \stackrel{(N_{a}\approx 1)\rightarrow (N_{a}\gg
1)}{\longrightarrow} \alpha_{a}(HUP)
\end{eqnarray}
in complete conformity to {\it Comment 2*.}

\subsection{Minimal Inverse Temperature and Measurability}

Now, let us return to the thermodynamic relation   (\ref{U12a}) in the case of equality:
\begin{equation}\label{U12a.new1}
\Delta \frac{1}{T}=\frac{k_B}{\Delta U} + \eta\,\Delta U,
\end{equation}
that is equivalent to the quadratic equation \begin{equation}\label{U12a.new2}
\eta\,(\Delta U)^{2}-\Delta\frac{1}{T}\Delta U+k_B=0.
\end{equation}
The discriminant of this equation, with due regard for formula (\ref{U12a*}), is equal to
\begin{equation}\label{U12a.new3}
\textit{D}=(\Delta\frac{1}{T})^{2}-4\eta
k_B=(\Delta\frac{1}{T})^{2}-4\alpha^{\prime}\frac{k^{2}_B}{E^{2}_p}\geq
0,
\end{equation}
leading directly to
$(\Delta\frac{1}{T})_{min}$
\begin{equation}\label{U12a.new4}
(\Delta\frac{1}{T})_{min}=2\surd \alpha^{\prime}\frac{k_B}{E_p}
\end{equation}
or, due to the fact that $k_B$ is constant, we have
\begin{equation}\label{U12a.new4*}
(\Delta\frac{1}{k_B T})_{min}=\frac{2\surd \alpha^{\prime}}{E_p}.
\end{equation}
It is clear that $(\Delta\frac{1}{T})_{min}$  corresponds to $T_{max}$
from formula (\ref{U15})
\begin{equation}\label{U12a.new5}
T_{max}\approx T_p\gg 0.
\end{equation}
In this case $\Delta\frac{1}{T}\approx \frac{1}{T}$ and, of course, we can assume that
\begin{equation}\label{U12a.new5}
(\frac{1}{T})_{min}\doteq \widetilde{\tau}=\frac{1}{T_{max}}.
\end{equation}
Trying to find from formula (\ref{U12a.new5}) a minimal unit of
measurability for the inverse temperature and introducing the
{``Integrality Condition'' (IC)} in line with the conditions
(\ref{Introd 2.4}),(\ref{Introd 2.4new})
\begin{equation}\label{U12a.new5*}
\frac{1}{T}=N_{1/T}\widetilde{\tau},
\end{equation}
where $N_{1/T}>0$ is  an integer number, we can introduce an
analog of the {\textbf{primary measurability} notion into
thermodynamics.
\\
\\{\bf Definition 3} ({\textbf{Primary Thermodynamic Measurability}})
\\(1) {\it  Let us define a quantity having the dimensions
of inverse temperature as {\bf primarily measurable} when it
satisfies the relation (\ref{U12a.new5*}).
\\(2)Let us define any physical quantity in thermodynamics
as {\bf primarily measurable} when its value is consistent with
point (1)  of this Definition}.
\\
\\{\bf Definition 3} in thermodynamics is analogous to the
{\textbf{Primary Measurability}} in a quantum theory ({\bf
Definition 1}).
\\Now we consider the quadratic equation (\ref{U12a.new2}) in
terms of  {\bf measurable quantities}  in the sense of {\bf
Definition
 3}. In accordance with this definition and
with formula (\ref{U12a.new5*}) $\Delta(1/T)$, we can write
\begin{equation}\label{U12a.new5**}
\Delta\frac{1}{T}=N_{\Delta(1/T)}\widetilde{\tau},
\end{equation}
where $N_{\Delta(1/T)}>0$ is  an integer number.
\\ The quadratic equation (\ref{U12a.new2}) takes the following form:
\begin{equation}\label{U12a.new2*}
\eta\,(\Delta U)^{2}-N_{\Delta(1/T)}\widetilde{\tau}\Delta U+k_B=0.
\end{equation}
Then, due to formula (\ref{U12a.new4*}), we can find the ''{\bf
measurable}'' roots of equation (\ref{U12a.new2*}) for $\Delta U$ as follows:
\begin{eqnarray}\label{U12a.new2*r}
(\Delta
U)_{meas,\pm}=\frac{[N_{\Delta(1/T)}\pm\sqrt{N_{\Delta(1/T)}^{2}-1}]\widetilde{\tau}}{2\eta}=\nonumber
\\=\frac{2k_B[N_{\Delta(1/T)}\pm\sqrt{N_{\Delta(1/T)}^{2}-1}]\widetilde{\tau}}{\widetilde{\tau}^{2}}=\nonumber
\\\frac{2k_B[N_{\Delta(1/T)}\pm\sqrt{N_{\Delta(1/T)}^{2}-1}]}{\widetilde{\tau}}.
\end{eqnarray}
The last line in (\ref{U12a.new2*r}) is associated with the obvious relation
$2\eta=\frac{\widetilde{\tau}^{2}}{2k_B}$.
\\ In this way we derive a complete analog of the corresponding relation
(\ref{root2}) from a quantum theory by replacement
\begin{eqnarray}\label{Change}
\Delta p_{\pm}\Rightarrow \Delta U_{meas,\pm};N_{\Delta
x}\Rightarrow N_{\Delta(1/T)};\hbar\Rightarrow k_B.
\end{eqnarray}
\\ As, for {\bf low temperatures and energies}, $T\ll T_{max}\propto
T_p$, we have $1/T\gg 1/T_p$ and hence $\Delta(1/T)\gg 1/T_p$ and
$N_{\Delta(1/T)}\gg 1$.
\\ Next, in analogy with Subsection 2.2, in formula (\ref{U12a.new2*r})
we can have only the minus-sign root, otherwise, at sufficiently
high $N_{\Delta(1/T)}\gg 1$ for $(\Delta U)_{meas,+}$ we can get
$(\Delta U)_{meas,+}\gg E_p$ . But this is impossible for low
temperatures (energies).
\\ On the contrary, the minus sign in (\ref{U12a.new2*r})
is consistent with high and low energies.
\\ So, taking the root value in (\ref{U12a.new2*r}) corresponding
to this sign and multiplying the nominator and denominator in
(\ref{U12a.new2*r}) by
$N_{\Delta(1/T)}+\sqrt{N_{\Delta(1/T)}^{2}-1}$,  we obtain
\begin{eqnarray}\label{U12a.new2*r-}
(\Delta
U)_{meas}=\frac{2k_B}{(N_{\Delta(1/T)}+\sqrt{N_{\Delta(1/T)}^{2}-1})\widetilde{\tau}}
\end{eqnarray}
to have a complete analog of the corresponding relation from
(\ref{root3}) in a quantum theory by substitution according to
formula (\ref{Change}).
\\Then it is clear that, in analogy with Subsection 2.2,
for low energies and temperatures $N_{\Delta(1/T)}\gg 1$
(\ref{U12a.new2*r-}) may be rewritten as
\begin{eqnarray}\label{U12a.new2*r-l}
(\Delta U)_{meas}\doteq(\Delta U)_{meas}(T\ll
T_{max})=\frac{2k_B}{(N_{\Delta(1/T)}+\sqrt{N_{\Delta(1/T)}^{2}-1})\widetilde{\tau}}\approx
\nonumber\\
\approx
\frac{k_B}{N_{\Delta(1/T)}\widetilde{\tau}},N_{\Delta(1/T)}\gg 1,
\end{eqnarray}
i.e. the Uncertainty Principle in Thermodynamics (UPT, formula
(\ref{U12})) is involved. In this case, due to the last formula,
$\Delta U_{meas}$ represents a {\bf primarily measurable}
thermodynamic quantity in the sense of {\bf Definition 3} to a
high accuracy.
\\ Of course, at high energies the last term in the formula
(\ref{U12a.new2*r-l}) is lacking and, for $T\approx
T_{max};N_{\Delta(1/T)}\approx 1$, we have:
\begin{eqnarray}\label{U12a.new2*r-h}
(\Delta U)_{meas}\doteq(\Delta U)_{meas}(T\approx T_{max})
=\frac{k_B}{1/2(N_{\Delta(1/T)}+\sqrt{N_{\Delta(1/T)}^{2}-1})\widetilde{\tau}},\nonumber\\
N_{\Delta(1/T)}\approx 1.
\end{eqnarray}
From (\ref{U12a.new2*r-h}) it follows that at high temperatures
(energies) $(\Delta U)_{meas}$  could hardly be a {\bf primarily
measurable} thermodynamic quantity. Because of this, it is
expedient to use a counterpart of {\bf Definition 2}.
\\
\\{\bf Definition 4. Generalized Measurability in Thermodynamics}
\\ Any physical quantity in thermodynamics may be referred to as {\bf
generalized-measurable} or, for simplicity, {\bf measurable} if
any of its values may be obtained in terms of the {{\textbf{Primary
Thermodynamic Measurability}} of {\bf Definition 3}.
\\
\\ In this way $(\Delta U)_{meas}$ from the formula
(\ref{U12a.new2*r-h}) is a {\bf measurable} quantity.
\\ Based on the preceding formulae, it is clear that we have the limiting transition
\begin{eqnarray}\label{P3Thermo}
(\Delta U)_{meas}(T\approx T_{max})
\stackrel{(N_{\Delta(1/T)}\approx 1)\rightarrow
(N_{\Delta(1/T)}\gg 1)}{\longrightarrow} (\Delta U)_{meas}(T\ll
T_{max}\propto T_p),
\end{eqnarray}
that is analogous to the corresponding formula (\ref{P3}) in a quantum theory.
\\ Therefore, in this case  the analog of {\it Comment
2*.} in Subsection 2.2 is valid.
\\{\it Comment 2* Thermodynamics}
\\{\it From the above formulae it follows that, within GUPT (\ref{U12a}),
the {\bf primarily measurable} variations (quantities)  are
derived, to a high accuracy, from the {\bf generalized-measurable}
variations (quantities) {\it only} in the low-temperature limit
$T\ll T_{max}\propto T_p$}.
\\ To conclude this Section, it seems logical to make several important
{\bf remarks}.
\\
\\{\bf R2.1} It is obvious that all the calculations associated with {\bf
measurability} of inverse temperature $\frac{1}{T}$  are valid for
$\beta=\frac{1}{k_B T}$ as well. Specifically, introducing
$\beta_{min}\doteq \widetilde{\beta}=\widetilde{\tau}/k_B$, we can
rewrite  all the corresponding formulae in the ''{\bf
measurable}'' variant with appropriate replacement.
\\
\\{\bf R2.2.} Naturally, the problem of compatibility between
the {\bf measurability} definitions in quantum theory  and  in
thermodynamics arises: is there any contradiction between  {\bf
Definition 1} from Subsection 2.2 and {\bf Definitions 3} from
Subsection 2.3 ?
\\ On the basis of the formulae (\ref{U15}) from Subsection 2.1
and (\ref{U12a.new5}) from Subsection 2.3 we can state: \\{\it
\textbf{measurability} in quantum theory and
\textbf{thermodynamic measurability} are completely compatible and
consistent as the minimal unit of inverse temperature
$\widetilde{\tau}$ is nothing else but the minimal time
$t_{min}=\tau$ up to a constant factor. And hence
$N_{1/T},(N_{\Delta(1/T)})$  is nothing else but $N_{t},(N_{\Delta
t})$ in (\ref{Introd 2.4new}). Then it is clear that
$N_{t}=N_{a=tc}$}.
\\
\\{\bf R2.3} Finally, from the above formulae (\ref{U12a.new2*r-l}),
(\ref{U12a.new2*r-h}) it follows that the  {\bf
measurable} temperature  $T$ is varying as follows:
\begin{eqnarray}\label{U12a.new2*--}
T=\frac{T_{max}}{N_{1/T}},T\ll T_{max}\propto T_p, N_{1/T}\gg 1;
\nonumber
\\T=\frac{T_{max}}{1/2(N_{1/T}+\sqrt{N_{1/T}^{2}-1})},T\approx T_{max}\propto T_p, N_{1/T}\approx
1.
\end{eqnarray}
In such a way {\bf measurable} temperature is a \textbf{discrete
quantity} but at low energies it is almost constantly varying,so
the theoretical calculations are very similar to those of the
well-known continuous theory.   In the reality, discreteness
manifests itself in the case of high energies only.

\section{Black Holes and Measurability}

Now let us show the applicability of the results from Section 2 to
a quantum theory of black holes. Consider the case of
Schwarzschild black hole. It seems logical to support the idea
suggested in the Introduction to the recent overview presented by
seven authors \cite{Hooft-2016}: ''Since for (asymptotically flat
Schwarzschild) black holes the temperatures increase as their
masses decrease, soon after Hawking’s discovery, it became clear
that a complete description of the evaporation process would
ultimately require a consistent quantum theory of gravity. This is
necessary as the semiclassical formulation of the emission process
breaks down during the final stages of the evaporation as
characterized by Planckian values of the temperature and spacetime
curvature''. Naturally, it is important to study the transition
from low to high energies in the indicated case.
\\ In this Section consideration is given to gravitational dynamics
at low $E\ll E_p$ and at high $E\approx E_p$ energies in the case
of the Schwarzschild black hole and in a more general case of the
space with static spherically-symmetric horizon in space-time in
terms of  {\bf
 measurable quantities} from the previous  Section.
\\ It should be noted that such spaces  and even considerably
 more general cases have been thoroughly studied from the viewpoint of gravitational
thermodynamics in remarkable works of professor T.Padmanbhan
\cite{Padm}--\cite{Padm13} (the list of references may be much
longer).
\\ First, the author has studied the above-mentioned case
in \cite{shalyt-IJMPD} and from the suggested viewpoint – in
\cite{Shalyt-AHEP2}. But, proceeding from Section 2 of the present
paper, it is possible to extend the results from
\cite{Shalyt-AHEP2}.
\\In what follows we use the symbols from \cite{Padm13}
which have been also used in \cite{Shalyt-AHEP2}. The case of a   static
spherically-symmetric horizon in space-time is considered,
 the  horizon being described by the metric
\begin{equation}\label{GT9}
ds^2 = -f(r) c^2 dt^2 + f^{-1}(r) dr^2 + r^2 d\Omega^2.
\end{equation}
The horizon location will be given by a simple zero of the
function $f(r)$, at the radius  $r=a$.
\\Then at the horizon $r=a$ Einstein's field equations (\cite{Padm13},
eq.(117)) take  the  form
\begin{equation}\label{GT11}
\frac{c^4}{G}\left[\frac{1}{2} f'(a)a - \frac{1}{2}\right] = 4\pi
P a^2
\end{equation}
where $P = T^{r}_{r}$ is the trace of the momentum-energy tensor
and radial pressure. Therewith, the condition $f(a)=0$ and
$f'(a)\ne 0$ must be fulfilled.
\\On the other hand, it is known that for horizon spaces one can introduce
the temperature that can be identified with an analytic
continuation to imaginary time. In the case under consideration
(\cite{Padm13}, eq.(116))
\begin{equation}\label{GT10}
k_BT=\frac{\hbar cf'(a)}{4\pi}.
\end{equation}
In \cite{Padm13} it is shown that in the initial (continuous) theory
the Einstein Equation for horizon spaces in the differential
form may be written as a thermodynamic identity (the first
principle of thermodynamics) (\cite{Padm13}, formula (119)):
\begin{equation}\label{GT12}
   \underbrace{\frac{{{\hbar}} cf'(a)}{4\pi}}_{\displaystyle{k_BT}}
    \ \underbrace{\frac{c^3}{G{{\hbar}}}d\left( \frac{1}{ 4} 4\pi a^2 \right)}_{
    \displaystyle{dS}}
  \ \underbrace{-\ \frac{1}{2}\frac{c^4 da}{G}}_{
    \displaystyle{-dE}}
 = \underbrace{P d \left( \frac{4\pi}{ 3}  a^3 \right)  }_{
    \displaystyle{P\, dV}},
\end{equation}
where, as noted above, $T$ -- temperature of the horizon surface,
$S$ --corresponding entropy, $E$-- internal energy, $V$ -- space
volume.
\\It is impossible to use (\ref{GT12}) in the formalism
under consideration because, as follows from the results given in
the previous section and in \cite{Shalyt-AHEP2},
  $da,dS,dE,dV$  are not {\bf measurable quantities}.
\\First, we assume that a value of the radius $r$ at the point $a$
is a  {\bf primarily measurable quantity} in the sense of {\bf
Definition 1} from Subsection 2.2., i.e. $a=a_{meas}=N_a\ell$,
where $N_a>0$ - integer, and the temperature $T$ from the
left-hand side of (\ref{GT10}) is the {\bf measurable} temperature
$T=T_{meas}$ in the sense  of  {\bf Definition 3} from Subsection
2.2.3.
\\Then, in terms of {\bf measurable} quantities,
first we can rewrite (\ref{GT11}) as
\begin{equation}\label{GT11.new1}
\frac{c^4}{G}\left[\frac{2\pi k_BT}{\hbar c}a_{meas} -
\frac{1}{2}\right] = 4\pi P a_{meas}^2.
\end{equation}
We express $a=a_{meas}$  in terms of the deformation parameter
$\alpha_{a}$ (formula (\ref{D1})) as
\begin{equation}\label{BH1}
a=\ell\alpha_{a}^{-1/2};
\end{equation}
the temperature $T$ is expressed in terms of $T_{max}\propto T_{p}$  from (\ref{U12a.new2*--}).
\\Then, considering that $T_p=E_p/k_B$, equation (\ref{GT11.new1}) may be given as
\begin{equation}\label{BH2}
\frac{c^4}{G}[\frac{\pi E_p}{\surd \alpha^{\prime}N_{1/T}\hbar
c}\ell\alpha_{a}^{1/2} - \frac{1}{2}\alpha_{a}] = 4\pi P \ell^2.
\end{equation}
Because $\ell=2\surd \alpha^{\prime}l_p$ and $l_p=\frac{\hbar
c}{E_p}$, we have
\begin{equation}\label{BH2.new}
\frac{c^4}{G}[\frac{2\pi E_p}{N_{1/T}\hbar c}l_p\alpha_{a}^{1/2} -
\frac{1}{2}\alpha_{a}] =\frac{c^4}{G}[\frac{2\pi
}{N_{1/T}}\alpha_{a}^{1/2} - \frac{1}{2}\alpha_{a}] = 4\pi P
\ell^2.
\end{equation}
Note that in its initial form \cite{Padm13} the equation
(\ref{GT11}) has been considered in a continuous theory, i.e. at
low energies $E\ll E_p$. Consequently, in the present formalism it
is implicitly meant that the ''measurable counterpart'' of
equation (\ref{GT11})  -- (\ref{GT11.new1}) (or the same
(\ref{BH2}),(\ref{BH2.new})) is also initially considered at low
energies, in particular, $N_{a} \gg 1,N_{1/T} \gg 1$.
\\Let us consider the possibility of generalizing
(\ref{BH2}),(\ref{BH2.new}) to high energies taking two different
cases.
\\
\\{\bf 3.1.} {\it Measurable case for low energies: $E\ll E_p$}.
Due to formula (\ref{root3.2}), $a=a_{meas}=N_a\ell$, where the
integer number is $N_a\gg 1$ or similarly $N_{1/T} \gg 1$. In this
case GUP, to a high accuracy,  is extended to HUP (formula
(\ref{root4}),(\ref{root5})).
\\As this takes place, $\alpha_{a}=\alpha_{a}(HUP)$  is a {\bf primarily
measurable} quantity ({\bf Definition 1}), $\alpha_{a}\approx
N_{a}^{-2}$, though taking a discrete series of values but varying
smoothly, in fact  {\it continuously}. (\ref{BH2}) is a quadratic
equation with respect to $\alpha_{a}^{1/2}\approx N_{a}^{-1}$
having the two parameters $N_{1/T}^{-1}$  and $P$. In this terms,
the equation (\ref{BH2.new})  may be rewritten as
\begin{equation}\label{BH2.new+}
\frac{c^4}{G}[\frac{2\pi }{N_{1/T}}\alpha_{a}^{1/2}(HUP) -
\frac{1}{2}\alpha_{a}(HUP)] = 4\pi P \ell^2.
\end{equation}
\\So, at low energies the equation (\ref{BH2.new}) (or
(\ref{BH2.new+})) written for the discretely-varying $\alpha_{a}$
may be  considered in a continuous theory.
\\As a result, in the case under study we can use the basic formulae
from a continuous theory considering them valid to a high
accuracy.
\\In particular, in the notation used for {\it Schwarzschild's   black
hole} \cite{Frolov}, we have
\begin{equation}\label{BH3}
r_s=N_a\ell=\frac{2GM}{c^{2}}; M=\frac{N_a\ell c^{2}}{2G}.
\end{equation}
As its temperature is given by the formula
\begin{equation}\label{BH3.1}
T_H=\frac{\hbar c^{3} }{8\pi GMk_B},
\end{equation}
at once we get
\begin{equation}\label{BH3.2}
T_H=\frac{\hbar c}{4\pi k_B N_a\ell}=\frac{\hbar c
\alpha_{a}^{1/2}}{4\pi k_B\ell}.
\end{equation}
Comparing this expression to the expression with high $N_{1/T}$
($N_{1/T}\gg 1$) for temperature from the equation
(\ref{U12a.new2*--}) that is involved in (\ref{GT11.new1}), we can
find that at low energies $E\ll E_p$, due to comment {\bf R2.2.}
from Subsection 2.3,  the number $N_{1/T}$ is actually coincident
with the number $N_a$:
\begin{equation}\label{BH4}
N_{1/T}=N_a=\alpha_{a}^{-1/2}(HUP).
\end{equation}
The substitution of the last expression from formula (\ref{BH3.2})
into the quadratic equation (\ref{BH2}) for $\alpha_{a}^{1/2}$
makes it a linear equation for $\alpha_{a}$ with a single
parameter $P$.
\\
\\{\bf 3.2.}{\it Measurable case for high energies:: $E\approx E_p$}.
Then, due to (\ref{root3.3}), $a$ is the {\bf generalized
measurable} quantity} $a=a_{meas}=1/2(N_{a}+\sqrt{N_{
a}^{2}-1})\ell$, with the integer $N_a \approx 1$.
\\ The quantity
\begin{equation}\label{BH4.corr}
\Delta a_{meas}(q)=1/2(N_{a}+\sqrt{N_{
a}^{2}-1})\ell-N_a\ell=1/2(\sqrt{N_{ a}^{2}-1}-N_{a})\ell
\end{equation}
may be considered as a {\bf quantum correction} for the {\bf
measurable} radius $r=a_{meas}$, that is infinitesimal at low
energies $E\ll E_p$  and not infinitesimal for high energies
$E\approx E_p$.
\\ In this case there is no possibility to replace GUP by HUP. In equation
(\ref{BH2}) $\alpha_{a}=\alpha_{a}(GUP)$  is a {\bf
generalized measurable} quantity ({\bf Definition 2}).
\\ As noted in formula (\ref{U12a.new2*--}) of Comment {\bf R2.3},
in this case the number  $N_{1/T}$  in equation (\ref{BH2.new})
is replaced by $1/2(N_{1/T}+\sqrt{N_{1/T}^{2}-1})$, i.e. the equation is of the form
\begin{equation}\label{BH2h}
\frac{c^4}{G}[\frac{2\pi
}{1/2(N_{1/T}+\sqrt{N_{1/T}^{2}-1})}\alpha_{a}^{1/2}(GUP) -
\frac{1}{2}\alpha_{a}(GUP)] = 4\pi P \ell^2.
\end{equation}
In so doing the theory becomes really discrete, and the solutions of
(\ref{BH2h}) take a discrete series of values for every
$N_a$ or ($\alpha_a(GUP)$) sufficiently close to
1.
\\ In this formalism for a ''quantum'' Schwarzschild black hole
(i.e. at high energies $E\approx
E_p$) formula (\ref{BH3.2}) is replaced by
\begin{equation}\label{BH3.2q}
T_H(Q)=\frac{\hbar c}{4\pi k_B
1/2(N_{a}+\sqrt{N_{a}^{2}-1})\ell}=\frac{\hbar c
\alpha_{a}^{1/2}(GUP)}{4\pi k_B\ell}.
\end{equation}
\\
\\ We should make several remarks which are important.
\\
\\{\bf Remark 3.3.}
\\ As noted in \cite{Shalyt-AHEP2}, the parameter $\alpha_a=\alpha_a(HUP)$,
within constant factors, is coincident with the Gaussian curvature
$K_{a}$ \cite{Geometry} corresponding to {\bf primary measurable}
$a=N_a\ell$:
\begin{equation}\label{GT16.C}
\alpha_a=\frac{\ell^{2}}{a^{2}}=\ell^{2}K_{a}.
\end{equation}
Because of this, the transition from $\alpha_a(HUP)$ to $\alpha_a(GUP)$ may be considered as a basis for
''quantum corrections''  to the Gaussian curvature
$K_{a}$ in the high-energy region $E\approx E_p$:
\begin{eqnarray}\label{BHGauss}
\alpha_{a}(GUP)-\alpha_{a}(HUP)=\ell^{2}[\frac{1}{1/4(N_{a}+\sqrt{N_{a}^{2}-1})^{2}\ell^{2}}-\frac{1}{N^{2}_a\ell^{2}}]=\nonumber\\
=\ell^{2}(K^{Q}_{a}-K_{a}),
\end{eqnarray}
where the ''measurable quantum Gaussian curvature '' $K^{Q}_{a}$
is defined as
\begin{eqnarray}\label{BHnew3}
K^{Q}_{a}\doteq
\frac{1}{1/4(N_{a}+\sqrt{N_{a}^{2}-1})^{2}\ell^{2}}.
\end{eqnarray}
In a similar way, with the use of formulae (\ref{BH3.2}) and
(\ref{BH3.1}), we can derive a ''measurable quantum correction ''
for the mass $M$ of a Schwarzschild black hole at high energies.
\\{\bf Remark 3.4.}
\\It is readily seen that a minimal value of   $N_a=1$ is {\it
unattainable} because in formula (\ref{root3.3})  we can obtain a
value of the length $l$ that is below the minimum $l<\ell$ for the
momenta and energies above the maximal ones, and that is
impossible. Thus, we always have $N_a\geq 2$. This fact was
indicated in \cite{shalyt2},\cite{shalyt3}, however, based on the
other approach.
\\
\\{\bf Remark 3.5.}
It is clear that we have the following transition:
\\
$$Eq.(\ref{BH2h})(E\approx E_{p})
\stackrel{(N_{a}\approx 1)\rightarrow (N_{a}\gg
1)}{\longrightarrow} Eq.(\ref{BH2.new+})(E\ll E_{p}).$$
\\{\bf Remark 3.6.}
So, all the members of the gravitational equation (\ref{BH2.new})
(and (\ref{BH2h}), respectively), apart from $P$, are expressed in
terms of the measurable parameter $\alpha_{a}$. From this it
follows that $P$ should be also expressed in terms of the
measurable parameter $\alpha_{a}$, i.e. $P=P(\alpha_{a})$: $E\ll
E_{p}$,  $P=P[\alpha_{a}(HUP)]$ at low energies and $E\approx
E_{p}$,$P=P[\alpha_{a}(GUP)]$ at high energies. Then, due to the
above formulae, we can have for a ''quantum'' Schwarzschild black
hole the “ horizon” gravitational equation in terms of {\bf
measurable} quantities
\begin{equation}\label{BH2Q}
(4\pi -1)\frac{c^4}{G}\alpha_{a}(GUP) = 8\pi P[\alpha_{a}(GUP)]
\ell^2,
\end{equation}
where $\alpha_{a}(GUP)$  takes a discrete series of the values
$\alpha_{a}(GUP)=(1/2(N_{a}+\sqrt{N_{a}^{2}-1}))^{-2}$; $N_{a}\geq
2$  is a small integer.

\section{Conclusion}

Taking a simple case as an example, in this paper the author has
successfully expressed almost all of the members in the
gravitational equation (excepting $P$) in terms of {\bf
measurable} quantities. In the general case the problem at hand is
as follows:
\\{\it the formulation of Gravity in terms of {\bf measurable}
quantities and also the derivation of a solution in terms of {\bf
measurable} quantities}.
\\Proceeding from the results obtained in \cite{Shalyt-AHEP2},
\cite{Shalyt-Entropy2016.1}, such a {\bf ''measurable''} Gravity
-- discrete theory that is practically continuous at low energies
$E\ll E_p$ and very close to the Einstein theory, though with some
principal differences.  By author’s opinion, in the low-energy
{\bf ''measurable''} variant of Gravity we should have no
solutions without physical meaning, specifically Godel's solution
\cite{Godel}.
\\ At high energies $E\approx E_p$ this {\bf
''measurable''} Gravity  should be really a discrete theory
enabling the transition to the low-energy {\bf ''measurable''}
variant of Gravity.
\\Still it is obvious that, to construct a {\bf measurable}
variant of Gravity at all the energy scales, in the {\it
general case} we need both the {\bf primarily measurable} variations
$\Delta p(N_{\Delta x}, HUP)$ (formula (\ref{root3.2})) and the {\bf
generalized-measurable} variations $\Delta p(N_{\Delta x}, GUP)$
from formula (\ref{root3.3}). The author believes that such
construction should be realized jointly with a construction of a
{\bf measurable} variant for Quantum Theory (QT).

\begin{center}
{\bf Conflict of Interests}
\end{center}
The author declares that there is no conflict of interests
regarding the publication of this paper.

\end{document}